# An Operational Approach to Quantum State Reduction

Masanao Ozawa*

*School of Informatics and Sciences, Nagoya University, Nagoya 464-01, Japan*

**Abstract**

An operational approach to quantum state reduction, the state change of the measured system caused by a measurement of an observable conditional upon the outcome of measurement, is founded without assuming the projection postulate in any stages of the measuring process. Whereas the conventional formula assumes that the probe measurement satisfies the projection postulate, a new formula for determining the state reduction shows that the state reduction does not depend on how the probe observable is measured, or in particular does not depend on whether the probe measurement satisfies the projection postulate or not, contrary to the longstanding attempts in showing how the macroscopic nature of probe detection provokes state reduction.

## 1. Introduction

Any measurement of an observable of a microscopic system changes the state of the measured system even in the case where the projection postulate may not hold. There are two ways of describing the state change caused by a measurement. One is *selective state change*, or so-called *state reduction*, which depends on the outcome of measurement. The other is *nonselective state change*, which does not depend on the outcome of measurement. Since the outcome of measurement is probabilistic in quantum mechanics, state reduction is also probabilistic. The conceptual status of state reduction, as probabilistic change of state, is still polemical in measurement theory [1], whereas the concept of nonselective state change is rather straightforward.

Any measuring process consists of two stages: the first stage is the interaction between the object and the apparatus, which transduces the measured observable to the probe observable, and the second stage is the detection of the probe observable, which amplifies the probe observable to a directly sensible macroscopic variable without further disturbing the object, or which is simply the measurement of the probe observable. Nonselective state change is well described by the open-system dynamics of the object system associated with the first stage. One of major polemical points concerning state reduction is whether the second stage plays any role in changing the state of the object dynamically.

---
*E-mail: e43252a@nucc.cc.nagoya-u.ac.jp

Consider a measurement of an observable $A$ of a quantum system $\mathbf{S}$ described by a Hilbert space $\mathcal{H}_\mathbf{S}$. Suppose that at the time of measurement the object is in the state (density operator) $\rho$, the apparatus is in the state $\sigma$, and the time evolution of the object-apparatus composite system during the interaction is represented by a unitary operator $U$ on the tensor product Hilbert space $\mathcal{H}_\mathbf{S} \otimes \mathcal{H}_\mathbf{A}$ where $\mathcal{H}_\mathbf{A}$ is the Hilbert space of the apparatus. Then the state $\rho'$ of the object just after measurement is obtained by the partial trace $\text{Tr}_\mathbf{A}$ over the Hilbert space $\mathcal{H}_\mathbf{A}$ of the apparatus as follows:

$$\rho' = \text{Tr}_\mathbf{A}[U(\rho \otimes \sigma)U^\dagger]. \tag{1}$$

This formula determines the nonselective state change $\rho \mapsto \rho'$. On the other hand, the state reduction, the probabilistic state change $\rho \mapsto \rho_a$ conditional upon the outcome $a$, is related to the nonselective state change by

$$\rho' = \sum_a P(a)\rho_a, \tag{2}$$

where $P(a)$, the probability of obtaining the outcome $a$, is determined by the statistical formula

$$P(a) = \text{Tr}[E^A(a)\rho], \tag{3}$$

where $E^A(a)$ is the projection operator with the range $\{\psi \in \mathcal{H}_\mathbf{S}|\ A\psi = a\psi\}$; if $a$ is an eigenvalue of $A$, $E^A(a)$ is the projection operator onto the eigenspace of $A$ corresponding to $a$, otherwise $E^A(a) = 0$.

One can naturally ask whether it is possible to determine uniquely the state reduction $\rho \mapsto \rho_a$ from the nonselective state change $\rho \mapsto \rho'$. The conventional approach stands in the negative. The conventional derivation of the state reduction from a given model of measuring process is to compute the state of the object-apparatus composite system just after the first stage assuming the Schrödinger equation for the composite system and to apply the projection postulate to the subsequent probe detection. Thus, the state $\rho_a$ is given by

$$\rho_a = \frac{\text{Tr}_\mathbf{A}[(1 \otimes E^M(a))U(\rho \otimes \sigma)U^\dagger(1 \otimes E^M(a))]}{\text{Tr}[(1 \otimes E^M(a))U(\rho \otimes \sigma)U^\dagger(1 \otimes E^M(a))]} \tag{4}$$

where $M$ is the probe observable [2].

The application of the projection postulate in the above derivation represents the state change in the second stage, which includes another *interaction* between the probe (i.e. the subsystem of the apparatus having the probe observable) and another part of the apparatus measuring the probe observable. Thus, the conventional derivation assigns the second stage part of the dynamical cause of the state reduction.

The validity of this derivation is, however, limited or questionable, apart from the interpretational questions such as "von Neumann's chain" argument [3], because of the following reasons:

1. The probe detection, such as photon counting, in some measuring apparatus does not satisfy the projection postulate [4]. In such a case, the conventional approach cannot determine the state reduction.

2. When the probe observable has continuous spectrum, the projection postulate to be applied cannot be formulated properly in the standard formulation of quantum mechanics [5]. Thus, the conventional approach does not apply to the state reduction for measurements of continuous observables.

3. When another measurement on the same object follows immediately after the first measurement, the measuring apparatus for the second measurement can interact with the object just after the first stage of the first measurement. The state reduction obtained by the conventional approach, which determines the state just after the second stage of the first measurement, cannot give the joint probability distribution of the outcomes of the above consecutive measurements [6].

In spite of the above points, it is usually claimed that state reduction has not yet occurred at the first stage but needs the further interaction between the apparatus and the observer's ego [7], between the apparatus and the environment [8], or between the probe and the macroscopic detector [9]—the application of the projection postulate to the object-apparatus composite system is considered to be an *ad hoc* expression of this kind of interaction. This claim is often supported by the following argument: The decomposition of the density operator $\rho'$ into $\rho_a$ with coefficient $P(a)$ in (2) is not mathematically unique. Hence, the partial trace formula (1) accompanied with the decomposition formula (2) does not determine the state reduction $\rho \mapsto \rho_a$. This argument is often summarized as the statement "the partial trace does not derive state reduction."

The purpose of this paper is to show that this argument is groudless. Contrary to the conventional view, we shall show that the state reduction $\rho \mapsto \rho_a$ can be derived by the nonselective state change $\rho \mapsto \rho'$ in the following way. Define state transformations $T$ and

$T_a$ by

$$T(\rho) = \rho', \tag{5}$$
$$T_a(\rho) = P(a)\rho_a. \tag{6}$$

Then, the statement that (2) holds for every initial state $\rho$ is equivalent to the relation

$$T = \sum_a T_a. \tag{7}$$

Thus, if we can show that the decomposition in (7) is unique, then we can determine $T_a$ from $T$ and hence the state reduction $\rho \mapsto \rho_a$ is derived by the relation

$$\rho_a = \frac{T_a(\rho)}{P(a)}. \tag{8}$$

If the transformation $T$ were to be decomposed into arbitrary mappings $T_a$ satisfying (6), this reformulation is equivalent to the original formulation in which the decomposition is not unique. The transformations $T_a$ are, nonetheless, not arbitrary but should satisfy certain general conditions naturally derived by the physical requirement for state transformations; indeed, we shall show that the transformations $T_a$ should be linear and positive transformations of the density operators. This paper will prove that under this physical requirement the decomposition in (7) is indeed unique. From this fact, it will be concluded that the claim "the partial trace does not derive state reduction" is groundless.

In this paper, we are confined to measurements of discrete observables. Hence, the word "observable" means "discrete observable" unless stated otherwise.

## 2. State change caused by measurement

In what follows we fix a measuring apparatus **A** measuring the observable $A$ of **S**. Suppose that one measures $A$ in the state $\rho(t)$ at the time $t$ using the apparatus **A**. If we denote by $\Pr\{A(t) = a \| \rho(t)\}$ the probability distribution of the outcome of this measurement, by the statistical formula we have

$$\Pr\{A(t) = a \| \rho(t)\} = \mathrm{Tr}[E^A(a)\rho(t)]. \tag{9}$$

Let $t + \Delta t$ be the time just after measurement. Then the measurement is carried out by the interaction between the object and the apparatus from the time $t$ to $t + \Delta t$, and the object is free from the apparatus after the time $t + \Delta t$. Let $\rho(t + \Delta t|a)$ be the state at $t + \Delta t$ of the object that leads to the outcome $a$. When $\Pr\{A(t) = a \| \rho(t)\} = 0$, the state $\rho(t + \Delta t|a)$ is not definite, and we let $\rho(t + \Delta t|a)$ be an arbitrarily chosen density operator for mathematical convenience.

In order to find the mathematical condition that characterizes the density operator $\rho(t + \Delta t|a)$, suppose that at the time $t + \Delta t$ the observer were to measure an arbitrary observable $X$ of the same object using an arbitrary apparatus measuring $X$. If we denote by $\Pr\{A(t) = a, X(t + \Delta t) = x \| \rho(t)\}$ the joint probability that the outcome of the $A$-measurement at $t$ is $a$ and that the outcome of the $X$-measurement at $t + \Delta t$ is $x$, then we have

$$\Pr\{A(t) = a, X(t + \Delta t) = x \| \rho(t)\} = \mathrm{Tr}[E^A(a)\rho(t)]\mathrm{Tr}[E^X(x)\rho(t + \Delta t|a)]. \tag{10}$$

When $\Pr\{A(t) = a \| \rho(t)\} \neq 0$, let $\Pr\{X(t + \Delta t) = x | A(t) = a \| \rho(t)\}$ be the conditional probability that the outcome of the $X$-measurement at $t + \Delta t$ is $x$ given that the outcome of the $A$-measurement at $t$ is $a$, i.e.,

$$\Pr\{X(t + \Delta t) = x | A(t) = a \| \rho(t)\} = \frac{\Pr\{A(t) = a, X(t + \Delta t) = x \| \rho(t)\}}{\Pr\{A(t) = a \| \rho(t)\}}. \tag{11}$$

Then, from (9) and (10) we have

$$\Pr\{X(t + \Delta t) = x | A(t) = a \| \rho(t)\} = \mathrm{Tr}[E^X(x)\rho(t + \Delta t|a)]. \tag{12}$$

Since $X$ is arbitrary, the density operator $\rho(t + \Delta t|a)$ satisfying (12) is uniquely determined. Conversely, we can derive also (10) from (12). Thus, when $\Pr\{A(t) = a \| \rho(t)\} \neq 0$, we can regard (12) as the mathematical definition of $\rho(t + \Delta t|a)$.

For any real number $a$, we define $T_a$ as the mapping that transforms any density operator $\rho$ to the trace class operator $\Pr\{A(t) = a \| \rho(t)\}\rho(t + \Delta t|a)$ when $\rho(t) = \rho$, i.e.,

$$T_a : \rho = \rho(t) \mapsto \Pr\{A(t) = a \| \rho(t)\}\rho(t + \Delta t|a). \tag{13}$$

It follows from (9), (10), and (13) that $T_a$ satisfies

$$\Pr\{A(t) = a, X(t + \Delta t) = x \| \rho\} = \mathrm{Tr}[E^X(x)T_a(\rho)]. \tag{14}$$

Suppose that the state $\rho\rho(t)$ is a mixture of the states $\rho_1$ and $\rho_2$, i.e.,

$$\rho = \alpha\rho_1 + (1 - \alpha)\rho_2 \tag{15}$$

where $0 < \alpha < 1$. This means that at the time $t$ the measured object **S** is sampled randomly from an ensemble of similar systems described by the density operator $\rho_1$ with probability $\alpha$ and from another ensemble described by the density operator $\rho_2$ with probability $1 - \alpha$. Thus we have naturally

$$\Pr\{A(t) = a, X(t + \Delta t) = x \| \rho\}$$
$$= \alpha \Pr\{A(t) = a, X(t + \Delta t) = x \| \rho_1\} + (1 - \alpha) \Pr\{A(t) = a, X(t + \Delta t) = x \| \rho_2\}. \quad (16)$$

From (14), (15), and the above equation, we have

$$\mathrm{Tr}\left[E^X(x)T_a\left(\alpha\rho_1 + (1-\alpha)\rho_2\right)\right] = \alpha\mathrm{Tr}[E^X(x)T_a(\rho_1)] + (1-\alpha)\mathrm{Tr}[E^X(x)T_a(\rho_2)]$$
$$= \mathrm{Tr}\left[E^X(x)\left(\alpha T_a(\rho_1) + (1-\alpha)T_a(\rho_2)\right)\right]. \quad (17)$$

Since $X$ is arbitrary, we have

$$T_a\left(\alpha\rho_1 + (1-\alpha)\rho_2\right) = \alpha T_a(\rho_1) + (1-\alpha)T_a(\rho_2). \quad (18)$$

Thus, $T_a$ is an affine transformation of the density operators, and hence it can be extended to a unique linear transformation of the trace class operators as follows. Every trace class operator $\sigma$ can be represented by four density operators $\sigma_1, \ldots, \sigma_4$ and four positive numbers $\lambda_1, \ldots, \lambda_4$ such as

$$\sigma = \lambda_1\sigma_1 - \lambda_2\sigma_2 + i\lambda_3\sigma_3 - i\lambda_4\sigma_4. \quad (19)$$

Then, the mapping $T_a$ can be extended to a unique linear transformation of any trace class operators by

$$T_a(\sigma) = \lambda_1 T_a(\sigma_1) - \lambda_2 T_a(\sigma_2) + i\lambda_3 T_a(\sigma_3) - i\lambda_4 T_a(\sigma_4). \quad (20)$$

For the proof, see [10].

We have now proved that for any apparatus **A** measuring $A$ there is uniquely a family $\{T_a | \ a \in \mathbf{R}\}$ of linear transformations of the trace class operators such that (14) holds for any observable $X$ and any density operator $\rho$. This family of linear transformations will be referred to as the *operational distribution* of the apparatus **A**.

## 3. Basic properties of the operational distribution

In this section, we shall examine the properties of the operational distribution of the apparatus **A**. An operator $X$ is said to be *positive*, written by $X \geq 0$, iff for any vector $\psi$ we have $\langle\psi|X|\psi\rangle \geq 0$. A linear transformation $L$ of the operators $X$ is said to be *positive* if we have $L(X) \geq 0$ whenever $X \geq 0$. For any vector $\psi$, choose $X$ and $x$ so that $E^X(x) = |\psi\rangle\langle\psi|$ and substitute them in (14), and by the positivity of the probability we have

$$\langle\psi|T_a(\rho)|\psi\rangle \geq 0$$

for any density operator $\rho$. Since every positive trace class operator is a positive multiple of a density operator, it is shown that $T_a$ is a positive linear transformation.

The following relations are obvious from (13):

$$\Pr\{A(t) = a \| \rho(t)\} = \operatorname{Tr}\left[T_a\left(\rho(t)\right)\right], \tag{21}$$

$$\rho(t + \Delta t | a) = \frac{T_a\left(\rho(t)\right)}{\operatorname{Tr}\left[T_a\left(\rho(t)\right)\right]}. \tag{22}$$

In (22), we have assumed $\Pr\{A(t) = a \| \rho(t)\} \neq 0$.

Since nonselective state change does not depend on the outcome of measurement, the state just after nonselective state change is statistically equivalent to the state of the system sampled randomly from the ensemble that is the mixture of systems in the state $\rho(t + \Delta t | a)$ with the relative frequency $\Pr\{A(t) = a \| \rho(t)\}$. Thus, the state, denoted by $\rho(t + \Delta t)$, of the system **S** yielded by the nonselective state change is given by

$$\rho(t + \Delta t) = \sum_a \Pr\{A(t) = a \| \rho(t)\} \rho(t + \Delta t | a). \tag{23}$$

Define the mapping $T$ by

$$T : \rho = \rho(t) \mapsto \rho(t + \Delta t) \tag{24}$$

for any density operator $\rho = \rho(t)$. Then, $T$ represents mathematically the nonselective state change caused by the apparatus **A**. From (13), (23), and (24), for any density operator $\rho$ we have

$$T(\rho) = \sum_a T_a(\rho). \tag{25}$$

Thus, by (18) for any density operators $\rho_1$ and $\rho_2$ and any $\alpha$ with $0 < \alpha < 1$ we have

$$T(\alpha \rho_1 + (1-\alpha)\rho_2) = \alpha T(\rho_1) + (1-\alpha) T(\rho_2). \tag{26}$$

It follows that, just as $T_a$, the mapping $T$ can be extended uniquely to a positive linear transformation of the trace class operators. By the extended $T$, equation (25) holds for any trace class operator $\rho$. Hence, $T$ satisfies

$$T = \sum_a T_a. \tag{27}$$

*Remark.* The rigorous treatment of the infinite sum in (27) is given in the following. Since $T_a = 0$ when $a$ is not an eigenvalue of $A$, the infinite sum in the right hand side is a countable sum. Let $\{a_1, a_2, \ldots\}$ be the set of eigenvalues of $A$. By (21) and by the countable additivity of probability, for any density operator $\rho$ we have

$$\lim_{k \to \infty} \left\| T(\rho) - \sum_{n=1}^{k} T_{a_n}(\rho) \right\|_\tau = 0,$$

where $\|\cdot\|_\tau$ denotes the trace norm. The convergence for any trace class operator $\rho$ is obvious from the decomposition as in (19) and (20).

By the above relation, $T$ is determined completely by the operational distribution $\{T_a|\ a \in \mathbf{R}\}$. This $T$ will be referred to as the *operation* of the apparatus $\mathbf{A}$. Since the operation $T$ transforms a density operator to a density operator, for any trace class operator $\rho$ we have

$$\mathrm{Tr}[T(\rho)] = \mathrm{Tr}[\rho]. \tag{28}$$

The corresponding characteristic property of the operational distribution is obtained from (9) and (21) as follows:

$$\mathrm{Tr}[T_a(\rho)] = \mathrm{Tr}[E^A(a)\rho] \tag{29}$$

for any real number $a$ and any trace class operator $\rho$.

Now we are ready to state the following important relations between operational distributions and operations. For the proof, see appendix A.

**Theorem 1.** *Let $\{T_a|\ a \in \mathbf{R}\}$ be the operational distribution of an arbitrary apparatus $\mathbf{A}$ measuring an observable $A$, and $T$ its operation. Then, for any real number $a$ and any trace class operator $\rho$ we have*

$$T_a(\rho) = T\left(E^A(a)\rho\right) = T\left(\rho E^A(a)\right) = T\left(E^A(a)\rho E^A(a)\right). \tag{30}$$

By the above theorem, the operational distribution $\{T_a|\ a \in \mathbf{R}\}$ of an arbitrary apparatus $\mathbf{A}$ measuring $A$ is determined uniquely by the operation $T$ of $\mathbf{A}$.

In view of the proof given in appendix A, the above theorem can be restated as the following statement.

**Theorem 2.** *If positive linear transformations $T$ and $\{T_a|\ a \in \mathbf{R}\}$ of the trace class operators satisfy (27)–(29), we have the relation (30).*

*Notes.* Mathematical theory of operational distributions was introduced by Davies and Lewis [11] based on the relations (27) and (28) as mathematical axioms; see also Davies [12]. Their relation with measuring processes was established in [5, 13] and applied to analyzing various measuring processes in [14].

## 4. State reduction and the dynamical description of measurement

In this section, we shall consider the relation between state reduction and the dynamical description of measurement.

Suppose that at the time $t$ of measurement the apparatus $\mathbf{A}$ is prepared in the state $\sigma$, and that the time evolution of the object-apparatus composite system from the time $t$ to $t+\Delta t$ is represented by a unitary operator $U$ on the tensor product Hilbert space $\mathcal{H}_\mathbf{S} \otimes \mathcal{H}_\mathbf{A}$. Then the composite system is in the state $U(\rho(t) \otimes \sigma)U^\dagger$ at the time $t+\Delta t$. Thus, the state $\rho(t+\Delta t)$ of the object at the time $t+\Delta t$ is given by

$$\rho(t+\Delta t) = \text{Tr}_\mathbf{A}[U(\rho(t) \otimes \sigma)U^\dagger]. \tag{31}$$

This formula determines the nonselective state change $\rho(t) \mapsto \rho(t+\Delta t)$. Thus, the operation $T$ of the apparatus $\mathbf{A}$ is determined by

$$T(\rho) = \text{Tr}_\mathbf{A}[U(\rho \otimes \sigma)U^\dagger] \tag{32}$$

for any trace class operator $\rho$. Therefore, according to Theorem 1 the operational distribution $\{T_a | \, a \in \mathbf{R}\}$ of the apparatus $\mathbf{A}$ is determined by

$$\begin{aligned} T_a(\rho) &= \text{Tr}_\mathbf{A}[U(E^A(a)\rho \otimes \sigma)U^\dagger] \\ &= \text{Tr}_\mathbf{A}[U(\rho E^A(a) \otimes \sigma)U^\dagger] \\ &= \text{Tr}_\mathbf{A}[U(E^A(a)\rho E^A(a) \otimes \sigma)U^\dagger]. \end{aligned} \tag{33}$$

It follows from (22) that when $\Pr\{A(t) = a \| \rho(t)\} \neq 0$ the state reduction $\rho(t) \mapsto \rho(t+\Delta t|a)$ is determined by

$$\begin{aligned} \rho(t+\Delta t|a) &= \frac{\text{Tr}_\mathbf{A}[U(E^A(a)\rho(t) \otimes \sigma)U^\dagger]}{\text{Tr}[E^A(a)\rho(t)]} \\ &= \frac{\text{Tr}_\mathbf{A}[U(\rho(t)E^A(a) \otimes \sigma)U^\dagger]}{\text{Tr}[E^A(a)\rho(t)]} \\ &= \frac{\text{Tr}_\mathbf{A}[U(E^A(a)\rho(t)E^A(a) \otimes \sigma)U^\dagger]}{\text{Tr}[E^A(a)\rho(t)]}. \end{aligned} \tag{34}$$

The above formulas show that the state reduction is determined only by what observable is measured, how the apparatus is prepared, and how the apparatus interacts with the object. Thus, the state reduction does not depend on how the probe observable is detected. In particular, it does not depend on whether the probe detection satisfies the projection postulate or not.

In the rest of this section, we shall compare the above formula, (34), with the conventional formula, (4). The conventional derivation — adopted, for instance, in [2, 15] — of the formula (4) runs as follows. The conventional approach also admits that the composite system is in the state $U(\rho(t) \otimes \sigma)U^\dagger$ at the time $t+\Delta t$. Then, it assumes that in the second stage of the measurement the probe observable $M$ is detected by the subsequent part of the measuring apparatus and that this detection statisfies the projection postulate. It is natural to assume that the probe observable $M$ has the same eigenvalues as the measured observable $A$ and that the outcome of the $M$-measurement is interpreted as the outcome of the $A$-measurement. Thus, by the projection postulate, if the probe detection leads to the outcome $a$, the object-probe composite system is in the state

$$\rho_{\mathbf{S}+\mathbf{A}}(t+\Delta t+\tau|a) = \frac{(1 \otimes E^M(a))U(\rho(t) \otimes \sigma)U^\dagger(1 \otimes E^M(a))}{\text{Tr}[(1 \otimes E^M(a))U(\rho(t) \otimes \sigma)U^\dagger(1 \otimes E^M)]} \tag{35}$$

at the time $t + \Delta t + \tau$, where $\tau$ is the time taken for the probe detection. It follows that the state $\rho(t + \Delta t + \tau|a)$ of the object leading to the outcome $a$ at the time $t + \Delta t + \tau$ is given by

$$\rho(t+\Delta t+\tau|a) = \frac{\text{Tr}_{\mathbf{A}}[(1 \otimes E^M(a))U(\rho(t) \otimes \sigma)U^\dagger(1 \otimes E^M(a))]}{\text{Tr}[(1 \otimes E^M(a))U(\rho(t) \otimes \sigma)U^\dagger(1 \otimes E^M)]}. \tag{36}$$

The state change $\rho(t) \to \rho(t + \Delta t + \tau|a)$ is what the conventional appraoch calls the state reduction; cf. (4). We have already discussed the conceptual difficulties in this derivation in Section 1. Two important points among them are that this derivation gives the state at the time just after the second stage — at the time $t + \Delta t + \tau$ — but not the first stage — at the time $t + \Delta t$ — and that this derivation holds only if the probe detection satisfies the projection postulate. Our new derivation circumvents the above difficulties and our new formula *removes all references to the probe detection*.

Now we shall show that our new formula still gives the *same* state transformation as the conventional approach. In order to see this, let us define state transformations $T'_a$ for any real $a$ by

$$T'_a(\rho) = \text{Tr}_{\mathbf{A}}[(1 \otimes E^M(a))U(\rho \otimes \sigma)U^\dagger(1 \otimes E^M(a))], \tag{37}$$

where $\rho$ is a trace class operator. Then, we have

$$\sum_a T'_a(\rho) = \text{Tr}_{\mathbf{A}}[U(\rho \otimes \sigma)U^\dagger] \tag{38}$$

for any trace class operator $\rho$ and hence

$$T = \sum_a T'_a. \tag{39}$$

The probe detection is naturally required to give the correct probability distribution of $A$ in the state $\rho(t) = \rho$ for any density operator $\rho$, i.e.,

$$\text{Tr}[(1 \otimes E^M(a))U(\rho(t) \otimes \sigma)U^\dagger] = \text{Tr}[E^A(a)\rho], \tag{40}$$

where Tr in the left-hand-side is taken over $\mathcal{H}_\mathbf{S} \otimes \mathcal{H}_\mathbf{A}$. Thus, we have

$$\text{Tr}[T'_a(\rho)] = \text{Tr}[E^A(a)\rho] \tag{41}$$

for any density operator $\rho$ and for all real $a$. It follows that $\{T'_a\}$ and $T$ satisfies the assumptions of Theorem 2, and hence we have

$$T'_a(\rho) = T(E^A(a)\rho) = T_a(\rho) \tag{42}$$

for all $\rho$ and $a$. Therefore, we have proved that, if the probe observable $M$ satisfying (40) is given, we have

$$\rho(t + \Delta t|a) = \frac{\text{Tr}_\mathbf{A}[(1 \otimes E^M(a))U(\rho(t) \otimes \sigma)U^\dagger(1 \otimes E^M(a))]}{\text{Tr}[(1 \otimes E^M(a))U(\rho(t) \otimes \sigma)U^\dagger(1 \otimes E^M)]} \tag{43}$$

for any outcome $a$ with $\Pr\{A(t) \neq 0 \| \rho(t)\}$. This shows that our argument also provides a new derivation of the conventional formula. New derivation does not assume the projection postulate for the probe detection and gives the state just after the first stage of the measurement. Therefore, our argument indeed enhance the validity of the conventional formula to the case where the probe detection does not satisfy the projection postulate and to the case where another measurement for the same object follows immediately after the first stage of the measurement.

## 5. Application to the measurement problem

In the conventional approach to the measurement problem, the discussion focuses on the measurement of an observable $A = \sum_n a_n |\phi_n\rangle\langle\phi_n|$ with nondegenerate eigenvalues that satisfies the projection postulate. In this case, the state reduction $\rho(t) \mapsto \rho(t + \Delta t|a_n)$ is determined by the projection postulate, or equivalently by the repeatability hypothesis, as follows:

$$\rho(t) \mapsto \rho(t + \Delta t|a_n) = |\phi_n\rangle\langle\phi_n|. \tag{44}$$

Hence, the operational distribution $\{T_a | a \in \mathbf{R}\}$ is determined by

$$T_a(\rho) = |\phi_n\rangle\langle\phi_n|\rho|\phi_n\rangle\langle\phi_n| \tag{45}$$

if $a = a_n$, otherwise $T_a(\rho) = 0$ for any trace class operator $\rho$. Accordingly, the operation $T$ is determined by

$$T(\rho) = \sum_n |\phi_n\rangle\langle\phi_n|\rho|\phi_n\rangle\langle\phi_n| \tag{46}$$

for any trace class operator $\rho$, and hence the nonselective state change $\rho(t) \mapsto \rho(t + \Delta t)$ is given by

$$\rho(t) \mapsto \rho(t + \Delta t) = \sum_n |\phi_n\rangle\langle\phi_n|\rho(t)|\phi_n\rangle\langle\phi_n|. \tag{47}$$

Suppose that the object **S** is in the state $\rho(t) = |\psi\rangle\langle\psi|$ at the time $t$ of measurement and that the apparatus is prepared in the state $\sigma = |\xi\rangle\langle\xi|$. Let $U$ be the unitary operator representing the time evolution of the object-apparatus composite system during the measurement. Then the composite system is in the state (vector) $U(\psi \otimes \xi)$ at $t + \Delta t$. It follows that the nonselective state change transforms the state of the object from $\rho(t)|\psi\rangle\langle\psi|$ to

$$\rho(t + \Delta t) = \text{Tr}_\mathbf{A}[U|\psi \otimes \xi\rangle\langle\psi \otimes \xi|U^\dagger]. \tag{48}$$

On the other hand, according to (46) the nonselective state change should satisfy

$$\rho(t + \Delta t) = \sum_n |\langle\phi_n|\psi\rangle|^2 |\phi_n\rangle\langle\phi_n|. \tag{49}$$

Thus, the first step to explain the state reduction is to find $U$ and $\xi$ satisfying

$$\text{Tr}_\mathbf{A}[U|\psi \otimes \xi\rangle\langle\psi \otimes \xi|U^\dagger] = \sum_n |\langle\phi_n|\psi\rangle|^2 |\phi_n\rangle\langle\phi_n|. \tag{50}$$

This problem is solved, for instance, by an arbitrary state vector $\xi$ and a unitary operator $U$ such that

$$U(\phi_n \otimes \xi) = \phi_n \otimes \xi_n \tag{51}$$

where $\{\xi_n\}$ is an arbitrary complete orthonormal basis [16].

The above $\xi$ and $U$ specify how the apparatus is prepared at the time of measurement and how it interacts with the object. In this way, the nonselective state change caused by the apparatus **A** is explained. Then, our question is as follows: Does the above argument with the state preparation $\xi$ and the interaction $U$ explains also the state reduction?

According to the conventional view, the answer is negative; although the state $U(\psi \otimes \xi)$ of the composite system leads to the nonselective state change $\rho(t) \mapsto \rho(t + \Delta t)$ by taking the partial trace in (48), the state reduction $|\psi\rangle\langle\psi| \mapsto |\phi_n\rangle\langle\phi_n|$, it is claimed, cannot be derived without further stage of measurement. The argument runs as follows. The partial trace formula (50) does not conclude that the object leading to the outcome $a_n$ is in the state $|\phi_n\rangle\langle\phi_n|$ just after measurement. Even though (50) might show that the state in the left hand side appears to be the mixture of the states $|\phi_n\rangle\langle\phi_n|$, such a decomposition of the density operator into the components of the mixture has no physical ground if another decomposition is mathematically possible. In fact, the decomposition of the density operator $\rho(t + \Delta t)$ into pure states with coefficients

$$\Pr\{A(t) = a_n \| \rho(t)\} = |\langle \phi_n | \psi \rangle|^2$$

is not unique in general. For example, if $k \neq l$ and

$$\Pr\{A(t) = a_k \| \rho(t)\} = \Pr\{A(t) = a_l \| \rho(t)\},$$

then, putting

$$\eta_n = 2^{-1/2}(\phi_k + \phi_l),$$
$$\eta_m = 2^{-1/2}(\phi_k - \phi_l),$$

we have another decomposition

$$\rho(t + \Delta t) = \sum_{n \neq k,l} |\langle \phi_n | \psi \rangle|^2 |\phi_n\rangle\langle\phi_n| + \sum_{n=k,l} |\langle \phi_n | \psi \rangle|^2 |\eta_n\rangle\langle\eta_n|.$$

Contrary to the above argument, according to Theorem 1, we can lead from (49) logically to the conclusion that the state $\rho(t + \Delta t | a_n)$ of object leading to the outcome $a_n$ is nothing but $|\phi_n\rangle\langle\phi_n|$. Our argument runs as follows. The decomposition (23) for any initial state $\rho(t)$ is mathematically equivalent to the decomposition (27) if $T_a$ should be an arbitrary mapping. Thus, the decomposition (27) might not be unique mathematically. Nonetheless, according to the physical requirement the operational distribution $\{T_a | a \in \mathbf{R}\}$ is not a family of arbitrary mappings but it should consist of positive linear transformations satisfying (29). Then, Theorem 1 shows that the decomposition (27) is unique under this physical requirement, and the operational distribution is determined by (33). In the present case, from (46) we have

$$T_{a_n}(\rho(t)) = T(E^A(a_n)|\psi\rangle\langle\psi|E^A(a_n))$$
$$= T(|\langle\phi_n|\psi\rangle|^2|\phi_n\rangle\langle\phi_n|)$$
$$= |\langle\phi_n|\psi\rangle|^2|\phi_n\rangle\langle\phi_n|,$$

and hence from (21) and (22) we conclude

$$\rho(t + \Delta t | a_n) = \frac{T_{a_n}(\rho(t))}{\Pr\{A(t) = a_n \| \rho\}}$$
$$= |\phi_n\rangle\langle\phi_n|.$$

Therefore, although the mathematical decomposition of (50) is not unique, the state reduction is derived from the unique *physical* decomposition of the partial trace formula (50).

## 6. Concluding remarks

This paper gives a new derivation of the state reduction from a given model of measuring apparatus. The new formula (34) determines the state reduction of a measurement of a given observable only from the unitary operator describing the time evolution of the object-apparatus composite system and the density operator describing the apparatus initial state without assuming the projection postulate in any stage of the measuring process. Thus, the state reduction is the same for any measurement of the same observable with the same interaction and the same apparatus preparation but it does not depend on the way of the probe detection. This concludes that the second stage of the measuring process has nothing to do with determining the state reduction as long as the probe observable is chosen properly.

This conclusion supports an experimental fact in quantum optics. Experimentalists in quantum optics usually calculate the state reduction along with the conventional method as if the probe detection might satisfy the projection postulate even in the case where the probe detection is carried out by photon counting for which the projection postulate does not hold. The reason why they reach the correct result under the incorrect assumption is now explained; according to the uniqueness of the operational distribution, the correct derivation of the state reduction always gives the same answer as the case where the probe detection satisfies the projection postulate. Thus, the current practice is justified.

In this paper, we are confined to measurements of discrete observables. The notion of state reduction caused by measurements of continuous observables involves much more mathematical complication. In the most general formulation of measurement models, the probability distribution of the outcome is described by normalized positive operator-valued measures and the state reduction is described by normalized completely positive map-valued measures (i.e. operational distributions in the general case). For the theory of quantum state reduction in this generality and its applications, we refer the reader to [5, 13, 14].

## Appendix

### A. The proof of Theorem 1

We shall first prove the following lemma.

**Lemma A.1.** *Let A be a discrete self-adjoint operator on a Hilbert space $\mathcal{H}$. Suppose that a positive linear transformation $T$ on the space $\tau c(\mathcal{H})$ of trace class operators on $\mathcal{H}$ and a family $\{T_a|\ a \in \mathbf{R}\}$ of positive linear transformations on $\tau c(\mathcal{H})$ satisfy the following conditions:*

1. $T = \sum_a T_a$.

2. $T^*(1) = 1$.

3. $T_a^*(1) = E^A(a)$ for any $a \in \mathbf{R}$.

In the above, $T^*$ and $T_a^*$ stand for the dual transformations of $T$ and $T_a$, respectively, on the space $\mathcal{L}(\mathcal{H})$ of bounded operators on $\mathcal{H}$. Then we have

$$T_a^*(X) = E^A(a)T^*(X) = T^*(X)E^A(a) = E^A(a)T^*(X)E^A(a) \tag{52}$$

for any $a \in \mathbf{R}$ and any $X \in \mathcal{L}(\mathcal{H})$.

*Proof.* By the linearity of $T_a^*$ and $T^*$ it suffices to show that (52) holds for any self-adjoint $X$. Let $X \in \mathcal{L}(\mathcal{H})$ be a bounded self-adjoint operator and $\{X\}''$ the commutative von Neumann algebra generated by $X$. Since the spectrum of $A$ is a countable set, we can order it as $\{a_1, a_2, \cdots\}$. The set $\mathcal{Z} = \ell^\infty[\{X\}'']$ of bounded sequences with values in $\{X\}''$ is a commutative von Neumann algebra with pointwise operations and the supremum norm. Since for any $\{L_n\} \in \mathcal{Z}$ the sequence $\{T_{a_n}^*(L_n)\}$ is a sequence of elements of $\mathcal{L}(\mathcal{H})$, we can define a mapping $\Phi$ from $\mathcal{Z}$ to $\mathcal{L}(\mathcal{H})$ by

$$\Phi : \{L_n\} \mapsto \sum_n T_{a_n}^*(L_n). \tag{53}$$

Note that for any density operator $\rho$ we have

$$\left| \mathrm{Tr}\left\{ \left[\sum_{n=1}^k T_{a_n}^*(L_n)\right] \rho \right\} \right| = \left| \sum_{n=1}^k \mathrm{Tr}[L_n T_{a_n}(\rho)] \right| \leq \sup_n \|L_n\| \, \mathrm{Tr}\left(\sum_n T_{a_n}(\rho)\right) = \sup_n \|L_n\|,$$

whence the sum in the right hand side of (53) converges in the weak operator topology. Since $\Phi$ is a positive linear mapping on a commutative C*-algebra, the Stinespring theorem [17] concludes that there exist a Hilbert space $\mathcal{W}$ Ca linear mapping $V : \mathcal{H} \to \mathcal{W}$ and a *-homomorphism $\pi : \mathcal{Z} \to \mathcal{L}(\mathcal{W})$ such that for any $\{L_n\}$ we have

$$\Phi(\{L_n\}) = V^\dagger \pi(\{L_n\}) V. \tag{54}$$

From condition 2, $\Phi$ preserves the units, and hence we have $V^\dagger V = 1$. Let $a = a_m$ and define $\{L_n'\}$ and $\{L_n''\} \in \mathcal{Z}$ by

$$\begin{aligned} L_n' &= \delta_{m,n} 1, \\ L_n'' &\equiv X. \end{aligned}$$

Then by conditions 1 and 3 we have

$$\Phi(\{L'_n\}) = E^A(a), \qquad (55)$$
$$\Phi(\{L''_n\}) = T^*(X), \qquad (56)$$
$$\Phi(\{L'_n\}\{L''_n\}) = T^*_a(X). \qquad (57)$$

From the relation $V^*V = 1$ and (55) we have

$$[\pi(\{L'_n\})V - VE^A(a)]^\dagger [\pi(\{L'_n\})V - VE^A(a)]$$
$$= V^\dagger \pi(\{L'_n\})\pi(\{L'_n\})V - V^\dagger \pi(\{L'_n\})VE^A(a) - E^A(a)V^\dagger \pi(\{L'_n\})V + E^A(a)V^\dagger VE^A(a)$$
$$= 0,$$

and consequently

$$\pi(\{L'_n\})V = VE^A(a) \qquad (58)$$
$$V^\dagger \pi(\{L'_n\}) = E^A(a)V^\dagger. \qquad (59)$$

Thus from (54)–(59) we have

$$T^*_a(X) = \Phi(\{L'_n\}\{L''_n\})$$
$$= V^\dagger \pi(\{L'_n\})\pi(\{L''_n\})V$$
$$= E^A(a)V^\dagger \pi(\{L''_n\})V$$
$$= E^A(a)\Phi(\{L''_n\})$$
$$= E^A(a)T^*(X).$$

Then, by the positivity of $T^*_a$, we have also

$$T^*_a(X) = T^*(X)E^A(a)$$

for any positive $X$, and hence the above relation holds for any $X \in \mathcal{L}(\mathcal{H})$ by the linearity of $T^*_a$. Thus, we have

$$T^*_a(X) = E^A(a)T^*(X) = T^*(X)E^A(a) = E^A(a)T^*(X)E^A(a).$$

This concludes the proof of (52). □

*The proof of Theorem 3.1.* Let $\{T_a|\ a \in \mathbf{R}\}$ be the operational distribution of the apparatus $\mathbf{A}$ and $T$ its operation. Then condition 1 of the above lemma holds by (27), condition 2 holds by (28), and condition 3 holds by (29). Let $X \in \mathcal{L}(\mathcal{H}_\mathbf{S})$. From the above lemma we have

$$\mathrm{Tr}[XT_a(\rho)] = \mathrm{Tr}[XT(\rho E^A(a))] = \mathrm{Tr}[XT(E^A(a)\rho)] = \mathrm{Tr}[XT(E^A(a)\rho E^A(a))]$$

for any trace class operator $\rho$. Since $X$ is arbitrary, (30) is concluded immediately. □